# GMCAO simulation tool development

Valentina Viotto[*ab], Elisa Portaluri[ab], Roberto Ragazzoni[ab], Carmelo Arcidiacono[bc], Maria Bergomi[ab], Marco Dima[ab], Jacopo Farinato[ab], Davide Greggio[ab], Demetrio Magrin[ab]

[a]INAF - Osservatorio Astronomico di Padova, Vicolo dell'osservatorio 5, Padova, Italia
[b]ADONI – ADaptive Optics National laboratory in Italy
[c]INAF – Osservatorio Astronomico di Bologna, Via Gobetti 93/3, Bologna, Italia

**ABSTRACT**

Global MCAO aims to exploit a very wide technical field of view to find AO-suitable NGSs, with the goal to increase the overall sky coverage. The concept foresees the use of numerical entities, called Virtual Deformable Mirrors, to deal with the nominally thin depth of focus reduction, due to the field of view size. These objects act together as a best fit of the atmospheric layers behavior, in a limited number of conjugation altitudes, so to become the starting point for a further optimization of the real deformable mirrors shapes for the correction of the -smaller- scientific field. We developed a simulator, which numerically combines, in a Layer-Oriented fashion, the measurements of a given number of wavefront sensors, each dedicated to one reference star, to optimize the performance in the NGSs directions.

Here we describe some details of the numerical code employed in the simulator, along with the philosophy behind some of the algorithms involved, listing the main goals and assumptions. Several details, including, for instance, how the number and conjugation heights of the VDMs are chosen in the simulation code, are briefly given. Furthermore, we also discuss the possible approaches to define a merit function to optimize the best solution. Finally, after an overview of the remaining issues and limitations of the method, numerical results obtained studying the influence of $C_n^2$ profiles on the reconstruction quality and the delivered SR in a number of fields in the sky are given.

## 1. INTRODUCTION

The main goal of the Global Multi-Conjugated Adaptive Optics (GMCAO)[1] concept is to extend the sky coverage to stars-poor directions in the sky, being these peculiar areas of great interest for science. This interest is not only related to a general lack on information, but in some cases, the peculiar darkness of these zones makes them appealing to specific science case, like high redshift galaxy surveys[2].

The approach envisaged by GMCAO to push the sky coverage to higher values is to look for Natural Guide Stars (NGS) in a technical Field of View (FoV) wider than the ones typically used in conventional Multi-Conjugated Adaptive Optics (MCAO[3]), as shown in Figure 1. This FoV extension is possible for the next generation Extremely Large Telescopes (ELT), characterized by a wide collecting area, which ensures a partial superposition of the footprints up to high altitude in the atmosphere, allowing to disentangle the altitude at which the detected turbulence is originated. On the other hand, the increase of the technical FoV, of course, leads to the consequent depth of focus reduction, which translates into the need for a higher number of correcting components in the AO loop. GMCAO overcomes this drawback numerically introducing a large number of Virtual Deformable Mirrors (VDM), which are then the starting point for an optimization of the real Deformable Mirrors (DM) shapes to allow the correction of a smaller scientific FoV. This kind of approach, which retrieves the information about the turbulence from sources positioned away from the areas to be actually corrected has been somehow proven with the existing MOAO[4] systems.

Another characteristic of the devised system, and resemblance with MOAO, is the fact that the WaveFront Sensors (WFS) will work in open loop mode. This is actually only partially true, since in the ELTs the AO system usually includes one of the main mirrors in the telescope optical train as the first correcting element, conjugated to the ground layer, and its correction will be experienced by the GMCAO WFSs too.

---

[*] valentina.viotto@oapd.inaf.it

The WFS proposed for GMCAO is called Very Linear WFS (VL-WFS)[5], and combines the sensitivity property of the pyramid WFS[6] with the linearity characteristics of a highly linear metrology system. To make the pyramid operate in its best regime[7][8][9], in fact, a further AO loop is closed with a local DM, inside each VL-WFS. New generation DMs, equipped with high precision capacitive sensors can, in principle, allow to avoid the mentioned metrology system.

Here we present the current status of a numerical simulator, developed in IDL language, which aims to derive the expected performance, in terms of Strehl Ratio (SR), of a GMCAO system, mounted on a given ELT.

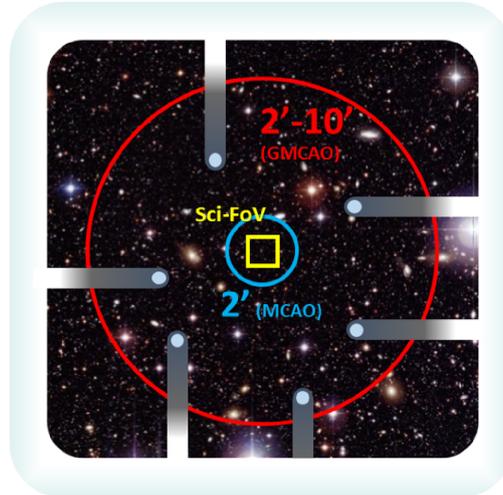

Figure 1 GMCAO Fields of View definitions

## 2. INPUT PARAMETERS

Our code accepts a number of input parameters, for which the final performance can be retrieved, including telescope and system characteristics, atmospheric assumptions and purely GMCAO parameters.

### 2.1 System description

The first parameters needed by the simulator are ones describing the telescope assumed as a collector and the AO system compensators characteristics. Summarizing them, we can start from the telescope collecting area and pupil shape, which can be given as an input, especially if it is characterized by exotic shape or segmentation. If the pupil shape is not given, the code will produce a circular default pupil. Another crucial parameter of the system is of course the (mean) scientific wavelength at which the SR will be estimated and the scientific FoV to be optimized. It is also foreseen the possibility to include a further FoV (larger than the scientific one and smaller than the technical one, used to select the NGSs), inside which the NGSs will be rejected by the system. This has been introduced to consider cases in which the GMCAO system needs to physically co-exist with other AO systems on the same telescope, e.g. an MCAO system, which uses a given FoV to look for NGSs for low-orders sensing. Finally, other important system parameters are number, conjugation altitudes and spatial samplings of the real DMs, which can be used as actual compensators in the AO loop.

### 2.2 Atmospheric model

An atmospheric model is needed to simulate the turbulent layers, which perturb the incoming wavefront. In our formulation, the code accepts as inputs the $C_n^2$ profile (e.g. in Figure 2), the Fried parameter (measured at zenith), the preferred turbulence description between Kolmogorov and Von Karman model, inner and outer scales.

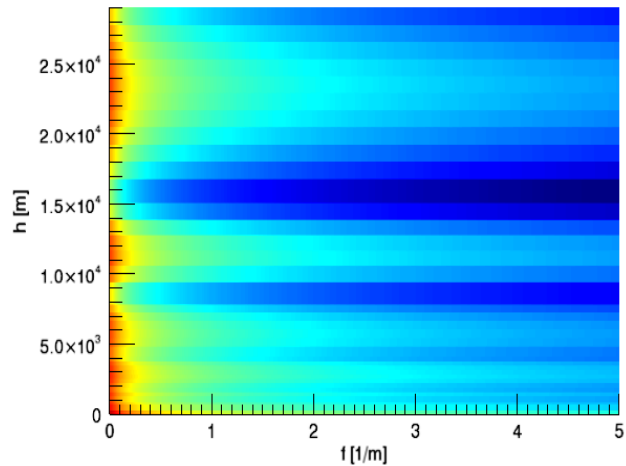

Figure 2 Example of input atmospheric $C_n^2$ profile distribution (in arbitrary units and logarithmic scale), for different frequencies.

## 2.3 Star field

Information on the NGSs position and magnitude is needed to retrieve the GMCAO performance. These input parameters can be either given manually, or computed in a circular asterism with a given radius, or derived from real star fields, retrieved from USNO-B1.0 catalogue, once the limiting magnitude is set. When many stars are available at the input coordinates, the star asterism selection optimizes the relative position of the stars so as to make the correction more uniform.

## 2.4 GMCAO parameters

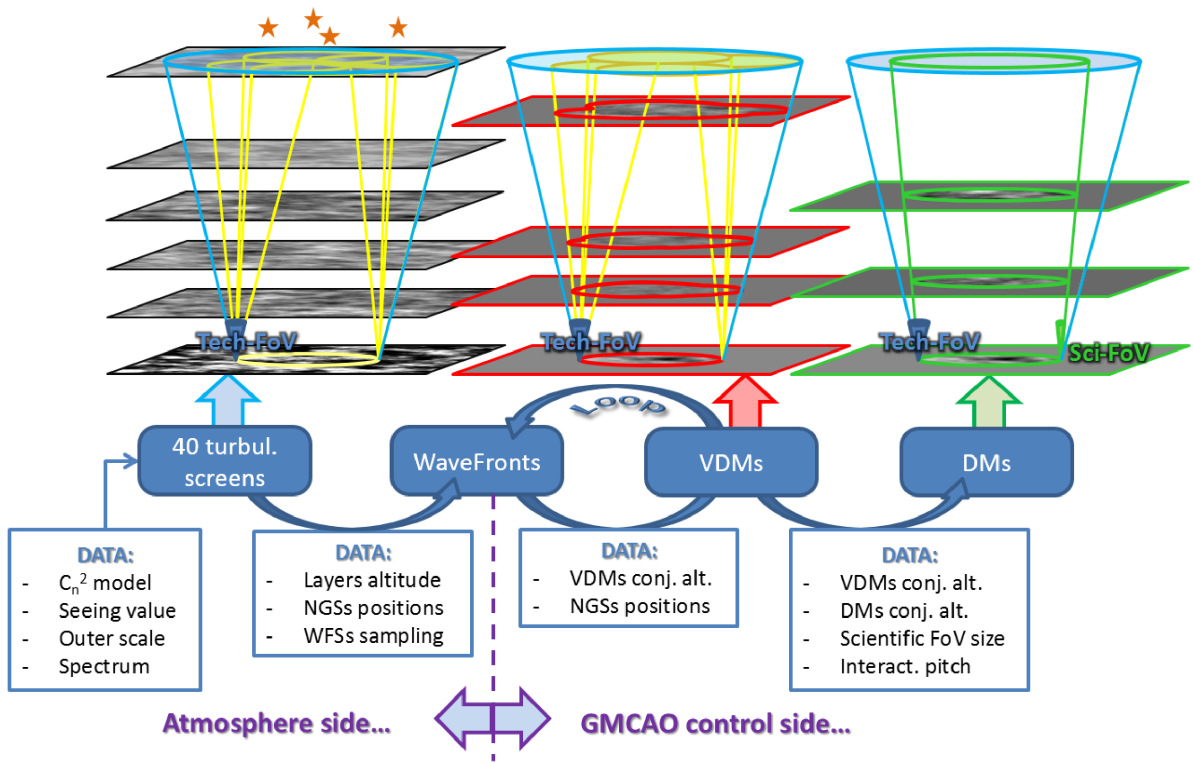

Figure 3 4-steps approach for the Tomographic Simulation Tool.

Concerning GMCAO parameters, we recall here that getting a high sky coverage is the primary goal of this technique, so the main parameters can be identified as the technical FoV size, which is directly related to the chance to find suitable references, and the limiting magnitude, driving the way in which the NGSs can be selected. Most of the input parameters (technical FoV, number of VDMs and their conjugation heights, number of probes, minimum distance between two NGSs). The last important input, needed to describe the GMCAO system, is the single WFS performance curve at the wavelength used for sensing, as a function of the source magnitude.

## 3. SIMULATION TOOL CODE FLUX

To compute a first order estimation of the behavior of the system performance in terms of delivered Strehl Ratio (SR) on the scientific FoV, we developed an IDL tool, reproducing the main aspects of the GMCAO system.

The GMCAO performance simulation tool is divided in a number of functions computing the different steps required by the overall simulation, the main of which are reported in Figure 4, which describes the logical flow of the information inside the code, with particular emphasis on the levels at which the inputs required to the user are used (only 1st time reported).

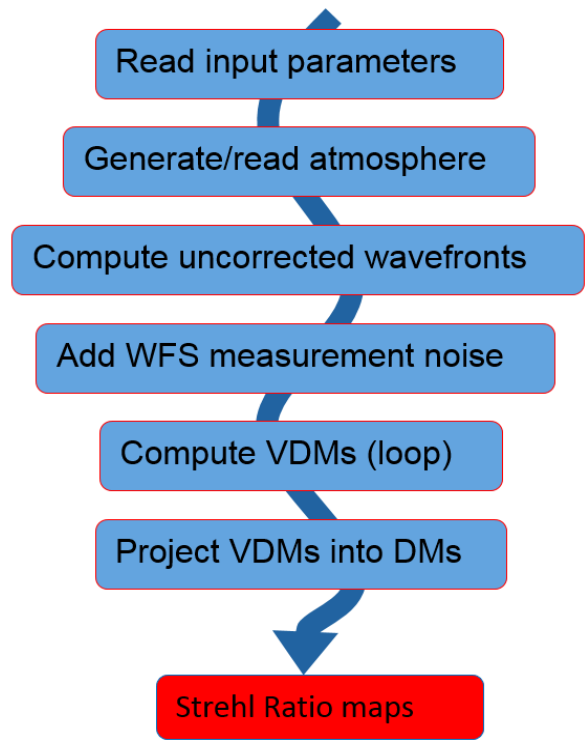

Figure 4 Simulation Tool logical flux.

### 3.1 Atmospheric layers computation

The code produces a synthetic frozen (i.e. not evolving) atmosphere, which obeys to the input $C_n^2$ model and seeing parameter. Each layer power spectrum follows a modified Von Karman slope, whose inner and outer scale can be set by the user, together with the spatial scale of the screen, produced as a 2D matrix. Different profiles can be used to compare the performance in different configurations. In the current issue of the code, the temporal evolution of the atmosphere is not simulated at this stage, but only taken into account in the response curve of each wavefront sensor, working in locally closed loop.

### 3.2 Wavefront computation

The shape of the perturbed wavefront, as it is retrieved by each VL-WFS, is obtained considering the asterism geometry and projecting the layers shape into the direction seen by the VL-WFS, when looking at its reference star in the sky. The geometry itself can be given as an input parameter to the system, or can be circular a-priori, with a given radius (in this case the position of the stars is slightly perturbed to avoid aliasing effect. For the layers aberration projection, sub-pixel shifts are considered, so not to be limited by the layers spatial sampling. After the uncorrected wavefront shape is computed, no more a-priori knowledge of the atmosphere is used.

To include the effect of different magnitudes of the reference stars, a VL-WFS response curve, in which the SR that can be obtained by the loop is given as a function of the guide star magnitude, is needed as an input. Following the input curve, a high frequency noise, whose minimum frequency corresponds to the local DM sampling, is associated to each star wavefront.

### 3.3 VDMs computation loop

The wavefronts retrieved by the VL-WFSs are the inputs for the VDMs computation. This is done inside a loop, in which the wavefronts shapes is assumed as a combination of the light distortion coming at each of the selected altitudes, in the same way. So, the VDMs are computed geometrically back projecting the wavefronts in the NGSs directions and applying a fraction of their shape to each of the VDMs conjugation altitudes. The "real" (this time is not an assumption, because we know where the VDMs lay) effect of the so-computed VDMs on the original wavefronts is then estimated, again, geometrically projecting them into the VL-WFSs directions. Figure 5 shows an example of retrieved high-altitude-conjugated VDM shape.

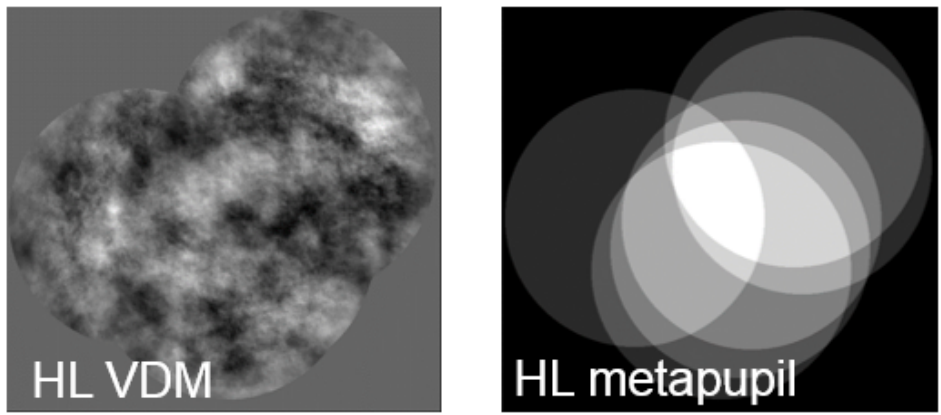

Figure 5 High layer VDM and metapupil shapes.

The residual aberration is then used as a starting point to start the loop again. After a certain number of iterations (limited by a limit threshold in the loop convergence and minimum residuals amplitude), the loop (example of loop convergence is shown in Figure 6) minimizes the difference between the actual measurements and the VDMs simulated projections onto the wavefront sensors. If the number of VDMs is too large with respect to the number of references, the problem becomes ill posed and there are a number of possible solutions, giving the same minimization level, meaning that it is possible, without any additional crosscheck, to obtain VDMs with odd shapes (they are not mimicking physically acceptable atmospheric layers, e.g. including discontinuities). An edge-smoothing function helps in avoiding this happens.

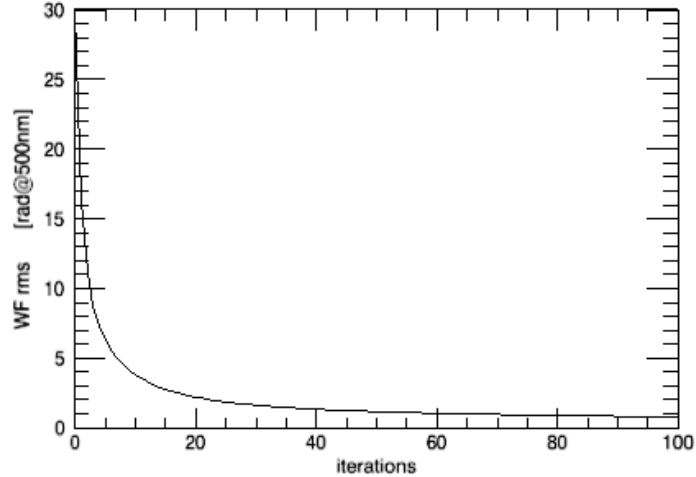

Figure 6 Example of residual wavefront rms convergence during VDMs loop.

### 3.4 DMs computation

The shape to be given to the real DMs is computed projecting the VDMs shapes with a gain, related to the distance of the conjugation altitudes of the VDM and the actual DM, and a smoothing function whose kernel represent the scientific FoV projection from the VDM to the DM conjugation altitude. In this way, the higher spatial scales, which are not common between the different directions inside the scientific FoV, are smeared out over the full metapupil and will not introduce noise in the reconstruction.

The problem of the correction of a FoV different from the one in which the references for the sensing have been selected is similar to the one faced by the MOAO systems, only the solution does not require for a full tomographic reconstruction, but can be simplified selecting the spatial frequencies, which are common to the whole FoV to be corrected.

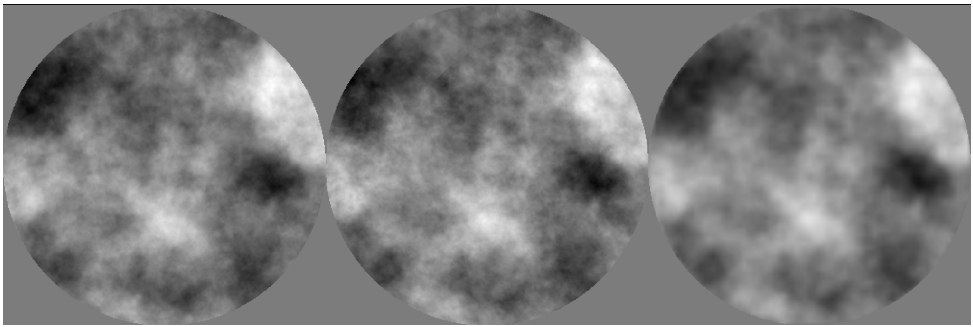

Figure 7 Examples, measured on-axis, of the sum of all input turbulence layers (left), VDMs (middle), and DMs (right). As expected, to allow a correction on the full scientific FoV, the DMs show to be effective up to a lower spatial frequency.

### 3.5 SR estimation

The code, after the correction, allows to estimate the SR which has been achieved, simply using the Marechal approximation on the scientific FoV wavefronts residuals. A SR map is then computed.

## 4. CONCLUSIONS

We implemented a code to estimate the performance of a GMCAO system, applied to a general telescope, and tested it with the E-ELT parameters, in different fields. Next step will be an extensive study of a GMCAO system sensitivity to asterism shapes and atmospheric profile, which, at the moment, give a strong indication of dependency, when testing a previous version of the code on fake random fields or real fields like the South Galactic Pole and the Chandra Deep Field

South (CDFS). Our previous estimation gave a SR up to 30%, with a mean value of 17% in the poorest star field analyzed (namely the CDFS), and this result has been used as an input for a study[2] on the suitability of the GMCAO approach for extra-galactic surveys purposes.

## REFERENCES

[1] Ragazzoni, R., Arcidiacono, C., Dima, M., Farinato, J., Magrin, D. and Viotto V., "Adaptive optics with solely natural guide stars for an extremely large telescope," in Proc. SPIE 7736, 23 (2010).

[2] Portaluri, E.; Viotto, V.; Ragazzoni, R.; Gullieuszik, M.; Bergomi, M.; Greggio, D.; Biondi, F.; Dima, M.; Magrin, D.; Farinato, J., "The Chandra Deep Field South as a test case for Global Multi Conjugate Adaptive Optics", MNRAS, 466, p.3569-3581 (2017)

[3] Beckers, J. M., "Increasing the size of the isoplanatic patch with multiconjugate adaptive optics", in Very Large Telescopes and their Instrumentation, ESO Conf., 2, 693 (1988)

[4] Hammer, F.; Puech, M.; Assemat, F. F.; Gendron, E.; Sayede, F.; Laporte, P.; Marteaud, M.; Liotard, A. and Zamkotsian, F., "FALCON: a concept to extend adaptive optics corrections to cosmological fields", in Proc. SPIE, 5382, 727 (2004)

[5] Magrin, D., Ragazzoni, R., Bergomi, M., Brunelli, A., Dima, M., Farinato, J. and Viotto, V., "Pyramid based locally closed loop wavefront sensor: an optomechanical study," Proc. Second AO4ELT Conf., 33 (2011)

[6] Ragazzoni, R., "Pupil plane wavefront sensing with an oscillating prism," Journal of Modern Optics 43, 289 (1996).

[7] Viotto, V.; Ragazzoni, R.; Bergomi, M.; Magrin, D.; Farinato, J., "Expected gain in the pyramid wavefront sensor with limited Strehl ratio", A&A, 593, A100 (2016)

[8] Ragazzoni, R. and Farinato, J., "Sensitivity of a pyramidic Wave Front sensor in closed loop Adaptive Optics," A&A 350, 23 (1999).

[9] Viotto, V., Magrin, D., Bergomi, M., Dima, M., Farinato, J., Marafatto, L. and Ragazzoni, R., "A study of Pyramid WFS behavior under imperfect illumination," Proc. Third AO4ELT Conf., 38 (2013)

[10] Viotto, V., Ragazzoni, R., Bergomi, M., Arcidiacono, C., Brunelli, A., Dima, M., Magrin, D., and Farinato, J., "Sky coverages on ELTs with a reference are much larger than the compensated one," Proc. Second AO4ELT Conf., 34 (2011).